\documentclass[twocolumn,showpacs,preprintnumbers,amsmath,amssymb]{revtex4}

\usepackage{graphicx}
\usepackage{dcolumn}
\usepackage{bm}

\begin{document}

\preprint{APS/123-QED}

\title{Molecular dynamics studies on spatial scale of low energy excitation in a simple polymer system}

\author{Akira Koyama}
 \affiliation{Toyota National College of Technology, Toyota, Aichi 471-8525, Japan}
 \email{koyama@toyota-ct.ac.jp}
\author{Takashi Odagaki}
 \affiliation{Department of Physics, Faculty of Science, Kyushu University, Fukuoka 812-8581, Japan}
\author{Koji Fukao}
 \affiliation{Department of Physics, Ritsumeikan University, Kusatsu, Shiga 525-8577, Japan}

\date{\today}

\begin{abstract}
A molecular dynamics simulation is performed to investigate spatial scale of
low energy excitation (LEE) in a single linear chain of united atoms.
The self part of the dynamic structure function, 
$S_\mathrm{S}(q,\omega)$, is obtained in a wide range in frequency space 
($\omega$) and reciprocal space ($q$).   
A broad peak corresponding to the LEE is detected at 
$\omega/2\pi=2.5 \times 10^{11} \ \mathrm{s^{-1}}$  ($\equiv \ \omega_{\mathrm{LEE}}/2\pi$) 
on the contour maps of $S_\mathrm{S}(q,\omega)$, near and below the  
glass transition temperature ($T_{\mathrm{g}}$=230 K).  
The $S_\mathrm{S}(q,\omega_{\mathrm{LEE}})$
is symmetric around a maximum along the logarithm of $q$.   
The inverse of $q_{\mathrm{max}}$, giving the maximum position of 
$S_\mathrm{S}(q,\omega_{\mathrm{LEE}})$, depends on temperature as
$2\pi/q_{\mathrm{max}}\sim T^{0.52}$ for $60 \ \mathrm{K}<T<T_{\mathrm{g}}$ and 
$2\pi/q_{\mathrm{max}}\sim T^{0.97}$ for $T_{\mathrm{g}}<T<600 \ \mathrm{K}$,
which is the spatial scale of the motion corresponding to the LEE at low temperatures.   
Based on a Gaussian approximation for the displacements of monomer groups 
which give rise to the motion relevant to the LEE, it is found that the 
number of monomers contained in a group is about 6. 

\end{abstract}

\maketitle

Low energy excitation (LEE) is observed universally in various glassy materials, 
such as amorphous metals \cite{Suc1983}, network glasses \cite{Buchenau1984,Gracia-Hernandez1993,Sokolov1993}, molecular glasses \cite{Dianoux1987,Bermejo1992,Yamamuro1997}, 
and glassy polymers \cite{Rosenberg1985,Inoue1991,Kanaya1993,Buchenau1994}.   
It has been studied intensively with the relation 
to anomalous thermal properties of the glassy materials in the temperature range from 10 to 30 K.

In many glassy materials, we can observe LEE as a broad peak 
in neutron and Raman scattering data at low energy range from 1 to 5 meV near 
and below the glass transition temperature ($T_\mathrm{g}$).   Because the temperature 
dependence of the peak height can be well fitted by Bose factor, 
it is often called as "Boson peak".   
LEE can be recognized as an excess excitation
superposed on that due to the vibrational density of state for a crystal (Debye solid), 
which is observed as an excess specific heat in the temperature range 
from 10 to 30 K \cite{Yamamuro1997}.   LEE affects the thermal conductivity in 
the temperature range, because the coupling between phonons carrying the heat 
and LEE occurs when their characteristic frequencies are close \cite{Zeller1971,Freemann1986,Buchenau1992}. 

So far, several models have been proposed for the elementary processes of LEE 
in order to explain the unusual behavior of the specific heat and thermal conductivity 
in the temperature range from 10 to 30 K \cite{Buchenau1992,Buchenau1991,Nakayama1999}.   Experimental researches
were performed to obtain the characteristic sizes of LEE for various materials \cite{Yamamuro1997,Tsukushi1998,Kanaya2000} 
using the proposed models.   
Although many results have been reported,  
the evidences justifying the models are not sufficient.

Recently, much attention has been paid to the origin of LEE, 
because significant improvements of experimental techniques enable us to 
examine the scattering intensity in a wide range of energy-momentum space \cite{5IDMRCS}.
It implies that we can in principle identify the characteristics of LEE directly 
without any specific models,
although the unified understanding of the origin of LEE has not yet been attained.

On the other hand, the energy and momentum ranges of LEE are accessible  
by molecular dynamics (MD) simulation.   Generally, MD simulation is 
a powerful tool to investigate atomic scale non-equilibrium phenomena.   
Thus, it is expected that the problems of LEE can successfully be clarified by MD simulation.   
Stimulated by the models proposed and the experimental results, MD simulations for 
various systems have been performed \cite{Habasaki1995,Schober1996,Roe1994,JainPablo2004},
where suggestive results have been reported.   


At the present stage, the information on the characteristics of LEE independent of the models
is highly required in order to elucidate the nature of LEE.
The purpose of this work is to 
examine the spatial scale of LEE by MD simulation.   We calculate self part of dynamic 
structure function which can be compared with experiments directly.
We choose a polymer model system to investigate LEE, because polymer 
is one of the common materials which easily vitrify, and its LEE has been studied extensively.

 We employ a united atom polyethylene (PE) model as the molecular model for the 
 simulations, which is essentially the same as that of our previous work \cite{Koyama2001}.   
 We consider energies of bond stretching, angle bending, and torsion of the model PE.   
 Truncated Lennard-Jones 12-6 potential is used to reproduce van der Waals interactions 
 between two united atoms apart more than three bonds, where a cut off radius is set to 
 1 nm.   The bulk polymer system is made of a single linear PE composed of 2000 united 
 atoms placed in an MD cell under three-dimensional periodic boundary conditions.   
 Newton's equations of motion are integrated by leap-flog algorithm with a time step 
 of 4 fs which is 1/20 of a period of the bond oscillation.   The temperature and the 
 pressure are controlled every 1 ps by ad hoc velocity rescaling method and loose-coupling 
 method \cite{Berendesen1984}, respectively, where the correction for the pressure due to the cut off 
 is taken into account.

An initial configuration of the model chain, with fixed bond lengths and bond 
angles, is generated using technique developed by Theodrou and Suter \cite{Theodorou1985}, and then a 
short MD simulation of 10 ps at a constant volume ($1.45 \ \ \mathrm{cm}^3 \mathrm{g}^{-1}$) and  
600 K is performed to reduce the overlaps of the atoms.   
This state is then fully relaxed for 1 ns at 1 atm and 600 K. 
The final state of 600 K obtained above is quenched to various 
temperatures between 200 K to 500 K, and the simulations of 2 ns are performed
at a constant pressure (1 atm) and the constant temperatures.  For the 
simulations at the temperatures below 200 K, the final state of 600 K is quenched to 
200 K, relaxed for 1 ns, and then quenched to the temperatures between 60 K to 180 K again.   
After the stepwise quenching, the simulations of 2 ns are carried out at a constant 
pressure (1 atm) and the temperatures.   We adopt the stepwise quenching procedure at 
the low temperatures to reproduce the temperature dependence of the specific volume 
of amorphous PE \cite{Koyama2001}.

Here we are interested in the LEE in the isotropic and homogeneous amorphous state of PE before 
it crystallizes, so that the total time of the simulation must be shorter than the 
characteristic time of the crystallization.   In the present study, the total simulation 
time of 2 ns at each temperature is much shorter than the time scale of the crystallization 
which is about the order of 10 ns observed previously \cite{Koyama2002,Koyama2003}.

In order to investigate the spatial scale of LEE, we evaluate self part of dynamic 
structure function, which can be measured by various scattering experiments.   
The self part of the dynamic structure function is defined as 
\begin{equation}
S_\mathrm{S}(\textbf{q},\omega)=\int_{-\infty}^{+\infty} F_\mathrm{S}(\textbf{q},t) \ \mathrm{e}^{-i \omega t} dt \ ,
\end{equation}
where \textbf{q} is the wave vector, $\omega$ is the angular frequency, \textit{t} is the time, and 
$F_\mathrm{S}(\textbf{q},t)$ 
is the self part of intermediate scattering function defined by 
\begin{equation}
F_\mathrm{S}(\textbf{q},t)=\frac{1}{N}<\sum_{j=1}^{N} \mathrm{e}^{-i\textbf{q}(\textbf{r}_{j}(t+t_0)-\textbf{r}_{j}(t_0))}>_{t_{0}}
\end{equation}
where \textit{N} is the number of the united atoms, $\textbf{r}_j (t)$ is the position vector 
of \textit{j}th united atom, and $<...>_{t_0}$ denotes the average about $t_0$.   
In our study, we employ an algorithm developed by Matsui et al. \cite{Matsui1994} to obtain 
$S_\mathrm{S}(\textbf{q},\omega)$,
where the 
$S_\mathrm{S}(\textbf{q},\omega)$
is calculated during the final 1 ns at a given temperature.   After calculating the 
$S_\mathrm{S}(\textbf{q},\omega)$,
we take an average of it about the solid angle of \textbf{q}, and then obtain 
$S_\mathrm{S}(q,\omega)$, where $q=|\textbf{q}|$.
We set the \textit{q} and $\omega$ ranges to be 
$5.3\times10^8 \ \mathrm{m}^{-1} < q/2\pi < 4.6\times10^{10} \ \mathrm{m}^{-1}$
and 
$9.5\times10^8 \ \mathrm{s}^{-1} < \omega/2\pi < 2.5\times10^{14} \ \mathrm{s}^{-1} \ 
(3.9\times10^{-3} \ \mathrm{meV} < \hbar\omega < 1.0\times10^3 \ \mathrm{meV})$, 
where $\hbar$ is the Planck constant divided by 2$\pi$.

\begin{figure}
\includegraphics[width=6cm]{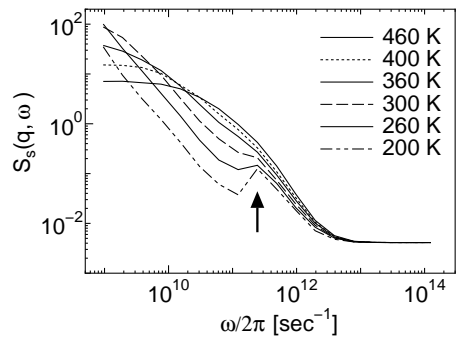}
\caption{
$S_\mathrm{S}(q,\omega)$ vs. $\omega/2\pi$. 
The results of the temperatures from 200 to 460 K are plotted. 
The arrow indicates the position of the LEE.
}
\end{figure}

In figure 1, $S_\mathrm{S}(q, \omega)$'s at different temperatures are plotted, 
where we choose the \textit{q} value corresponding to the first-peak position of 
static structure function (structure factor) at each temperature.   We recognize 
a single relaxation above 400 K.   Here, we call the 
relaxation as the alpha process.   At 360 K, a shoulder appears around $2\times10^{11}\ \mathrm{s}^{-1}$, 
and the single relaxation at the higher temperatures seems to be splitted into two 
deferent relaxation processes.   The relaxation located in the 
lower frequency region is the alpha process.   As the temperature decreases, 
the alpha process shifts toward the low $\omega$ region, and the shoulder gradually changes 
into a peak which corresponds to the LEE.   The maximum position of $2.5\times10^{11}\ \mathrm{s}^{-1}
(=\omega_\mathrm{LEE}/2\pi)$ corresponds roughly to that of LEE reported previously in the literature
 \cite{Inoue1991,Kanaya1993}.   
It has also been reported that the $\omega_\mathrm{LEE}$ does not depend on temperature generally, 
while the relaxation time of alpha process depends strongly on temperature.   Such characteristics are 
well reproduced in Fig 1.

\begin{figure*}
\includegraphics[width=10cm]{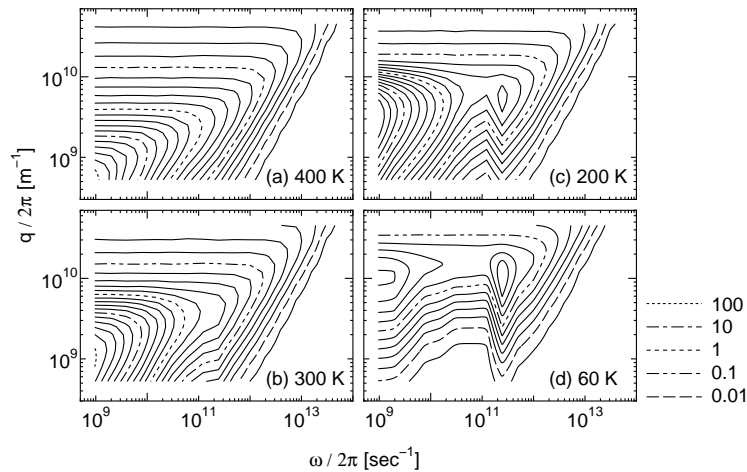}
\caption{
Contour maps of $\log S_\mathrm{S}(q,\omega)$ as a function of logarithm 
of $\omega$ and $q$ at the temperatures of 400 K (a), 300 K (b), 200 K (c), and 60 K (d).
}
\end{figure*}

To know the overall profile of the $S_\mathrm{S}(q,\omega)$, the contour maps of 
$\log[S_\mathrm{S}(q,\omega)]$ as a function of the logarithm of $q/2\pi$ and $\omega/2\pi$ 
are shown in Fig.2.   At 400 K, a single ridge can be recognized from the upper-right corner 
to the lower-left one in Fig.2(a).
As the temperature decreases, the part of the ridge below $\omega/2\pi=2\times10^{11} \ \mathrm{s}^{-1}$ 
shifts toward higher \textit{q}. 
At the same time, another ridge appears almost along the line of $\omega/2\pi=2\times10^{11} \ \mathrm{s}^{-1}$,
and hence we can observe that the single ridge at 400 K is splitted into two different ones at 300 K.
This second ridge at $\omega/2\pi=2\times10^{11} \ \mathrm{s}^{-1}$ in Fig.2(b) corresponds to the shoulder
observed in Fig.1. At 200 K, the original ridge below $\omega/2\pi=2\times10^{11} \ \mathrm{s}^{-1}$ shifts 
toward higher $q$ further, and the second ridge becomes much more distinct, so that a peak can be observed at 
$\omega=\omega_{\mathrm{LEE}}$ and $q/2\pi=5\times10^{9} \ \mathrm{m}^{-1}$. With decreasing temperature further,
the peak position shifts toward higher $q$, while it does not move along the horizontal axis.
The peak position along the vertical axis corresponds to the spatial scale of the LEE at a given temperature.

In Figure 3(a), the $S_\mathrm{S}(q,\omega_\mathrm{LEE})$ is plotted as a function of 
the logarithm of $q/2\pi$, in order to examine the temperature dependence of the spatial 
scale of the LEE.   We can notice that all the profiles at the temperatures are symmetrical 
around maxima.   The $S_\mathrm{S}(q,\omega_\mathrm{LEE})$'s can be well fitted by an 
empirical function of $X(q)=A/\{(q/q_\mathrm{max})^B +(q/q_\mathrm{max})^{-B} \}$+constant, 
where $q_\mathrm{max}$, $A$, and $B$ are the position, the height, and the shape factor of the 
peak, respectively.   
Figure 3 (b) shows the log-log plot of the $2\pi/q_\mathrm{max}$ vs. the temperature.   
The $2\pi/q_\mathrm{max}$ decreases with decreasing temperature. 
Non-bonded inter monomer distance (NBIMD) which is estimated from the first peak position
of the static structure function also decreases from $5.2\times10^{-10} \ \mathrm{m}$ 
to $4.5\times10^{-10} \ \mathrm{m}$ with decreasing temperature (not shown). 
The $2\pi/q_\mathrm{max}$ at 600 K is comparable to the the NBIMD.
But, it decreases further beyond the range of the NBIMD.
At 60 K, the $2\pi/q_\mathrm{max}$ is about $1.0\times10^{-10} \ \mathrm{m}$ which is smaller 
than a covalent bond length of $1.54\times10^{-10} \ \mathrm{m}$.
We notice that two straight lines intersect at 230 K corresponding to the $T_\mathrm{g}$ defined 
in our previous work \cite{Koyama2001}, which means that the two different power laws hold above and below $T_\mathrm{g}$.   
After fitting the observed data in Fig.3(b) to a function of $2\pi/q_\mathrm{max}=C\times T^{\beta}$, 
we obtained $\beta=0.52$ and $C=1.1\times10^{-11}\ \mathrm{m}\ \mathrm{K}^{-\beta}$ for $60\ \mathrm{K}<T<T_\mathrm{g}$, 
and $\beta=0.97$ and $C=9.3\times10^{-13}\ \mathrm{m}\ \mathrm{K}^{-\beta}$ for $T_\mathrm{g}<T<600\ \mathrm{K}$.

Here, we discuss the properties of the LEE observed in our results.   
Below $T_\mathrm{g}$, the LEE has a major contribution to the molecular motion,
because the characteristic time of the alpha process is very large compared to that of the LEE.   
Now, we assume that monomer groups, giving rise to the motion associated with the LEE, are trapped 
in potential energy minima.   It is expected that the monomer groups oscillate around their 
average positions, because large rearrangement of the monomers are prohibited below $T_\mathrm{g}$.   
If the displacements of the mass centers of the monomer groups from their average positions 
do not correlate and they obey Gaussian distribution, the root mean square of the displacement
should be proportional to $T^{1/2}$.    
Since the the temperature dependence of the $2\pi/q_\mathrm{max}$ which can be regarded as the averaged 
amplitude of the oscillation of the monomer groups is almost consistent with the Gaussian approximation, 
we conclude that the motion corresponding to the LEE should be described by the above scenario 
in the temperature range of $60\ \mathrm{K}<T<T_\mathrm{g}$, 

As the temperature increases, the characteristic time of the alpha process becomes smaller.   
Then the contributions of the LEE and the alpha process to the value of $2\pi/q_\mathrm{max}$ become 
superposed in the temperature range above $T_\mathrm{g}$, so that the temperature dependence 
of $2\pi/q_\mathrm{max}$ becomes strong.   We can consider that the monomer groups are no longer trapped at 
fixed positions.   The mass centers of the groups are drifted by the alpha process.   
It is reasonable that the crossover point of the power law appears at $T_\mathrm{g}$. 

\begin{figure}
\includegraphics[width=6cm]{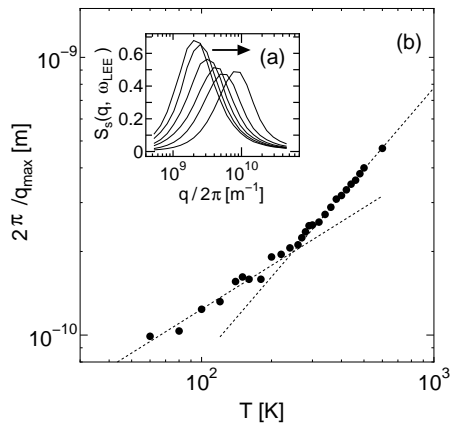}
\caption{
$S_\mathrm{S}(q,\omega_\mathrm{LEE})$ vs. $q/2\pi$ (a), and $2\pi/q_\mathrm{max}$ vs. temperature (b).  
The results of $S_\mathrm{S}(q,\omega_\mathrm{LEE})$ from 100 K to 600 K are plotted with an interval of 100 K. 
The maximum position of the peak of $S_\mathrm{S}(q,\omega_\mathrm{LEE})$ shifts along to the arrow 
with decreasing temperature. The broken lines in (b) are the fitting lines, 
where the values of the slopes above and below $T_\mathrm{g}$ is 0.97 and 0.52 respectively.
}
\end{figure}

According to the Gaussian approximation for the displacements of the mass centers of 
the monomer groups below $T_\mathrm{g}$, we estimate the number of the monomers contained in the group.   
In this Gaussian approximation, square of the averaged amplitude of the oscillation is derived as 
$(2\pi/q_\mathrm{max})^2=3k_\mathrm{B}T/m_\mathrm{eff}\omega_\mathrm{LEE}^2$, where $m_\mathrm{eff}$ 
is the effective mass of the monomer groups.    Substituting $2\pi/q_\mathrm{max} = C\times T^{1/2}$, 
the effective mass is written as $m_\mathrm{eff} =3k_\mathrm{B}/C^2\omega_\mathrm{LEE}^2$.   
From the fitting result of $C$ for the data of $60\ \mathrm{K}<T<T_\mathrm{g}$ in Fig 3(b), 
we obtained $m_\mathrm{eff} =82 \ \mathrm{g\ mol^{-1}}$ which corresponds to 5.9 monomers 
for our model PE.   It is smaller than the known characteristic sizes concerning to LEE \cite{Kanaya2000}.  
However, our method to derive it is much simpler and more direct than the others.

In summary, we have performed the MD simulation to investigate the LEE of the model PE 
system in the temperature range above and below $T_\mathrm{g}$.   We have succeeded 
in observing the LEE at $\omega_\mathrm{LEE}/2\pi=2.5\times10^{11}\ \mathrm{s}^{-1}$ 
in the contour map of $S_\mathrm{S}(q,\omega)$.   The $S_\mathrm{S}(q,\omega_\mathrm{LEE})$ 
as a function of logarithm of $q$ is symmetric around the maximum position of $q_\mathrm{max}$ 
at a given temperature.   The $2\pi/q_\mathrm{max}$, which corresponds to the spatial 
scale of the LEE at the low temperatures, depends on temperature as $2\pi/q_\mathrm{max}\sim T^{0.52}$ 
for $T<T_\mathrm{g}$ and $2\pi/q_\mathrm{max}\sim T^{0.97}$ for $T>T_\mathrm{g}$.   Based on 
the Gaussian approximation for the displacements of the monomer groups which give rise 
to the motion regarding the LEE, it is found that the number of the monomers included 
in a group is 5.9.   
The problems of the LEE are not completely solved, but detailed analysis 
is under progress to clarify it, which will be published in a subsequent paper. 

The authors wish to express their gratitude to associate professor H. Takatsu of 
Toyota National College of Technology for his helpful proposal on our study.
This work was supported by a Grant-in-Aid for Scientific Research
(B) (No.15340139 and No.16340122) from Japan Society for the Promotion of Science.
 
\bibliography{PaperIV}

\end{document}